\documentstyle[prb,preprint,aps]{revtex}
\begin{document}
\draft
%
%
\title{Simple theory for spin-lattice relaxation in metallic rare earth
ferromagnets}
\author{W. H\"ubner and K. H. Bennemann}
\address{Institute for Theoretical Physics, Freie Universit\"at
Berlin, Arnimallee 14, D-14195 Berlin, Germany}
\date{\today}
\maketitle
\begin{abstract}
The spin-lattice relaxation time $\tau_{SL}$ is a key quantity both
for the dynamical response of ferromagnets excited by laser pulses and
as the speed limit of magneto-optical recording. Extending the
theory for the electron paramagnetic resonance of magnetic impurities to
spin-lattice relaxation in ferromagnetic rare earths we
calculate $\tau_{SL}$ for Gd and find a value of 48 ps in very good
agreement with time-resolved spin-polarized photoemission experiments. We
argue that the time scale for $\tau_{SL}$ in metals is essentially given
by the spin-orbit induced magnetocrystalline anisotropy energy.
\end{abstract}
\newpage
\noindent
{\bf I. INTRODUCTION}\\
The spin-lattice-relaxation
time $\tau_{SL}$ is a sensitive fingerprint for the strength
of the dynamical coupling between the spin system and the lattice.
This time is therefore of interest for the long-time spin response of magnetic
materials upon pulse laser lattice excitation. In a
ferromagnetic solid, this time is required to establish a new equilibrium
magnetization after a sudden change of the lattice temperature.
Thus, $\tau_{SL}$ is a key quantity for magneto-optical recording, since it
determines the maximum speed for magneto-optical Curie-point
writing~\cite{leo}.
In their pioneering work on Gd, Vaterlaus {\em et al.}~\cite{vaterlaus}
were the first to measure $\tau_{SL}$ in real-time
using time-resolved spin-polarized photoemission. This experiment was
performed with the pump and probe technique applying strong 10 ns laser
heating pulses followed by 60 ps weak probe pulses with variable delay
and yielded the result
\[
\tau_{SL} \;= \;(100 \;\pm \; 80) \;{\rm ps} \;\cong \; 5 \;- \;50 {\rm GHz}\;.
\]
This corresponds
to a gain by two or more orders of magnitude in speed compared to the
present state of the art of data processing, which is still rapidly improving.

Up to now there exists no calculation or theoretical explanation of this
result.
Thus, it is the goal of this paper to provide a theoretical approach to
spin-lattice relaxation in metallic rare-earth ferromagnets, which is based on
rate equations and an electronic model structure. We present a theory which,
despite of its simplicity, exhibits already the microscopic features of
spin-lattice relaxation in these materials.

The spin-lattice relaxation time $\tau_{SL}$ describes the time
required by the spins to reach thermal equilibrium with the lattice.
The lattice then operates as a heat bath if one neglects
the ``phonon bottleneck'' thus assuming perfect coupling to the external
environment via the phonons. Hereby the originally cold spins are flipped by
the phonons, and spin and phonon systems approach a common thermal equilibrium.
This is microscopically accomplished as follows: {\em The
spins couple to the anisotropic fluctuations of the crystal fields produced by
the phonons. This coupling is mediated by spin-orbit interaction.}
During this process neither a modification of the geometrical structure nor a
change of the magnetic phase (long range order) has to take place.

A typical scenario of the processes leading to
spin-lattice relaxation is a four-step process: (i) The laser beam hits
the sample and creates electron-hole pair excitations within 10$^{-15}$ sec.
(ii) The electronic system equilibrates at elevated temperatures by
electron-electron interactions within 10 fs. Note, the lattice is not yet
involved. The spin and charge dynamics at the elevated temperature
may already lead to the breakdown of magnetic long range order
in the case of intense fs laser pulses but not for ns heating, since
the electronic system is always close to equilibrium for ns photoemission.
(iii) The equilibrated electronic excitations decay via phonon cascades
within 10$^{-13}\;\ldots $ 10$^{-12}$ s
and heat up the phonon system, i. e. the lattice.
(iv) The phonons and the spin system reach their common equilibrium within
the spin lattice relaxation time $\tau_{SL}$ of 10$^{-10}$ s.
This is also the time which allows for the recovery of magnetism in many cases,
since the electronic equilibrium temperature after step (ii) might be much
larger than the Curie temperature,
whereas the common equilibrium temperature reached
after step (iv) for spins and phonon system is usually much lower
than the electronic equilibrium temperature and may also
be smaller than the Curie temperature. The characteristic {\em interactions}
of these four processes taking place on distinct time scales are:
(i) \mbox{\bf p$\;\cdot\;$A}, where {\bf p}
is the crystal momentum of the electrons and {\bf A} is
the vector potential of the laser photons, (ii) electron-electron
Coulomb interaction leading to dynamical charge and spin fluctuations,
(iii) electron-phonon interaction,
(iv) phonon-magnon interaction caused by {\em spin-orbit interaction} which we
will approximate by the static magnetocrystalline
anisotropy energy (see below).

The experiment by Vaterlaus {\em et al.} was the
first to measure the time evolution of the magnetic nonequilibrium
state on a {\em picosecond} time scale. Therefore, in this paper,
we exclusively address the long-time (ps to ns) response via the lattice
to compare with the above experiment.

To support the above scenario of a laser pulse causing on the ps timescale and
at not too low temperatures mainly the heating up of phonons, we compare the
specific heat of phonons, spins, and electrons~\cite{ashcroft} and find the
following:
The spins start to dominate the phonons at temperatures
\begin{equation}
T\;\leq \;T_{0} \;\approx \;0.1\;\Theta_{D}
\end{equation}
for fields
of about 1 Tesla (in the case of paramagnetic impurities). The electrons
start to dominate the phonons at temperatures
\begin{equation}
T\;\leq \;T_{0} \;\approx \;0.01\;\Theta_{D}\;.
\end{equation}
Typical Debye temperatures $\Theta_{D}$ in ferromagnets are 420 K (Fe),
385 K (Co), 375 K (Ni), 152 K (Gd), 186 K (Dy). (For details of these estimates
see appendix A). This crude
estimate shows already, that phonons are dominant at sufficiently long time
scales and not too low temperatures
(the experiment has been performed typically at temperatures between 30
and 300 K), whereas spins (corresponding to
Coulomb-correlated electrons) and finally electrons
(single-particle excitations) are going to
take over for lower temperatures but also for shorter times (100 fs and
shorter). Thus, it appears reasonable to focus on the phonons, since the
experiment has been done on the ps to ns time scale and at low temperatures.
However, it becomes immediately obvious that very interesting dynamical
properties of the electrons and spins are to be expected in faster (fs) pump
and
probe experiments which will definitely be available in the near future.
However, for this time window, the notion of spin-{\em lattice}
relaxation makes no sense any more, since electrons rather than phonons are
involved.\\
\\
{\bf II. MICROSCOPIC CALCULATION OF THE SPIN-LATTICE RELAXATION
TIME $\tau_{SL}$}\\
To calculate the spin-lattice relaxation
time $\tau_{SL}$ we start from the theoretical approaches successfully
applied to electron spin resonance (ESR) more
than three decades ago for magnetic impurities
embedded in a nonmagnetic host lattice
and adapt this treatment to the solid combining
phenomenological nonequilibrium thermodynamics (kinetic theory) and microscopic
equilibrium theory. Three processes (all involving phonons) contribute to
spin-lattice relaxation: The direct process (Fig. 1 (a), see appendix B),
the Orbach process~\cite{orbach} (Fig. 1 (b), see appendix B), both of them
being relevant only at very low temperatures, and the
Raman process (Fig. 1 (c)) which we consider here:
This process consists of a {\em spin-flip},
the {\em absorption} of a phonon of frequency $\omega $,
and of the {\em emission} of a phonon of
frequency $\omega\;+\;\omega_{0}$. The longitudinal relaxation rate $T_{1}$ in
this case is independent of the magnetic field~\cite{scott} and is given by
\begin{equation}
\frac{1}{T_{1}}\;\sim \;T^{7}\;\ldots \;T^{9}.
\end{equation}
The Raman process is a two-phonon process
of higher order which essentially uses the complete phonon spectrum.
This process dominates the Orbach process (and thus also the
direct process, see appendix B) for
\begin{equation}
\frac{\Delta _{1}}{k _{B}}\;\geq \;\Theta _{D}\;,
\end{equation}
where $\Delta _{1}$ is the crystal field splitting and $k _{B}$
is Boltzmann's constant. Nickel, for example, has
\begin{equation}
\frac{\Delta _{1}}{k _{B}}\;\approx \;688K\;\gg \;\Theta _{D}\;\approx
\;375K\;.
\end{equation}
Thus, for not too low temperatures, the
Raman-process is dominant for the spin-lattice relaxation rate.

Therefore, in view of the experimental conditions,
it appears justified to focus on
{\em Raman} determined spin-lattice relaxation in the solid which
should be valid at intermediate lattice temperatures and ps time scales. The
temperature range of validity forms probably the best compromise
between too large temperatures where the lattice becomes unstable (above
the melting point) or magnetism breaks down (above the Curie temperature)
and too low temperatures where direct and Orbach processes determine
the phonon induced relaxation or the phonons become
frozen. Besides, the Raman process is independent of the magnetic field.

Note that purely electronic mechanisms such as spin fluctuations in strongly
correlated electronic systems mediated by nuclear
spin-flips (for energy and angular momentum conservation) via hyperfine
interaction require even longer time scales
and are unimportant in this context since they do not involve the lattice.

To calculate now Raman-induced spin-lattice relaxation
in ferromagnetic rare earth solids we start from the theory for
spin-lattice relaxation in magnetic impurities~\cite{scott}. First we
consider the number of phonons in the volume $V$ and energy interval
[$\delta $,$\delta \;+\;d\delta $]
\begin{equation}
\rho (\delta )d\delta \;=\;\frac{3V\delta ^{2}d\delta }{2\pi ^{2}\hbar
^{3}v_{s}^{3}}\;,
\end{equation}
where $v_{s}$ is the speed of sound in the material
(e. g. Gd). The thermal occupation is given by the Bose factor:
\begin{equation}
\bar{p}_{0}(\delta )\;=\;\frac{1}{e^{\frac{\delta }{k_{B}T}}-1}.
\end{equation}
For the interaction, the usual crystal field expansion up to second
order in terms of the randomly fluctuating strains is used
\begin{equation}
H_{c}^{\prime \prime } \;\approx \;\varepsilon _{1}\varepsilon
_{2}\sum_{mn}v_{n}^{m},
\end{equation}
since the Raman effect is of second order (see Fig. 1(c)).
The transition probability from state
$|b >$ to $|a >$ is then given by
\begin{equation}
w_{b\rightarrow a}\;=\;\int\frac{2\pi }{\hbar }\mid<b,\bar{p}_{0}(\delta _{1}),
\bar{p}_{0}(\delta _{2})\mid H_{c}^{\prime \prime }\mid a,
\bar{p}_{0}(\delta _{1})-1,\bar{p}_{0}(\delta _{2})+1
>\mid ^{2}\rho (\delta _{2})\rho (\delta _{1})d\delta _{1}.
\end{equation}
Including the
processes of stimulated emission, absorption, and spontaneous emission
the rate equation for the change of the occupation numbers of the
levels $|b >$ and $|a >$ is given by ($\rho $ is the mass density of the solid)
\begin{equation}
\dot{N_{b}}=-N_{b}w_{b\rightarrow a}+N_{a}w_{a\rightarrow b}
=-\dot{N_{a}}=K[-N_{b}\bar{p}_{0}(\delta )-N_{b}+N_{a}\bar{p}_{0}(\delta )].
\end{equation}
Using eqs. (8) and (9) this leads to
\begin{equation}
\dot{N_{b}}=\frac{9\sum_{mn}\mid<a\mid v_{n}^{m}\mid b>
\mid ^{2}}{8 \rho ^{2}\pi^{3}\hbar^{7}v_{s}^{10}}\int[N_{a}\bar{p}_{0}
(\delta _{2})[\bar{p}_{0}(\delta _{1})+1]-N_{b}\bar{p}_{0}(\delta _{1})
[\bar{p}_{0}(\delta _{2})+1]]\delta _{1}^{6}d\delta _{1}.
\end{equation}
Here, it has been used that the
square of the matrix elements of the strains $\varepsilon $ assumes the value
\begin{equation}
\frac{\delta[\bar{p}_{0}(\delta )+1]}{2Mv_{s}^{2}}\;,
\end{equation}
where $M$ is the crystal mass. Using the plausible assumptions
\begin{equation}
\delta\;\ll \;k_{B}T,\;\delta\;\ll \;\delta_{1}
\end{equation}
and the abbreviations
\begin{equation}
n\;=\;N_{a}\;-\;N_{b},\;N\;=\;N_{a}\;+\;N_{b}\;\;{\rm and}
\;\;n_{0}\;=\;N\tanh(
\frac{\delta }{2k_{B}T})
\end{equation}
yields the kinetic equation of spin-lattice relaxation
\begin{equation}
\dot{n}\;=\;-\frac{1}{\tau _{SL,Raman}}(n\;-\;n_{0})\;.
\end{equation}
The microscopic calculation of the spin-lattice relaxation rate
(which is the kinetic coefficient of the rate equation) gives then the result
\begin{equation}
\frac{1}{\tau _{SL,Raman}}\;=\;\frac{9\sum_{mn}\mid<a\mid v_{n}^{m}\mid b>
\mid ^{2}}{8 \rho ^{2}\pi^{3}\hbar^{7}v_{s}^{10}}\int_{0}^{k_{B}\Theta _{D}}
\frac{\delta _{1}^{6}e^{\frac{\delta _{1}}{k_{B}T}}d\delta _{1}}
{(e^{\frac{\delta _{1}}{k_{B}T}}-1)^{2}}.
\end{equation}
Using our previous estimate for the
magnetocrystalline anisotropy, which is discussed in some detail in appendix C,
\begin{equation}
\sum_{mn}\mid<a\mid v_{n}^{m}\mid b>\mid ^{2}\;=\;\mid E_{anisotropy}\mid ^{2}
\;=\;\mid 735\mu eV\mid ^{2}\;,
\end{equation}
this microscopic
theory finally yields for the spin-lattice relaxation time in Gd a value of
\begin{equation}
\tau _{SL,Raman}\;=\;48 ps\;.
\end{equation}
This result is in excellent agreement with the experimental value of (100 $\pm$
80) ps. The main issue here is that obviously the energy scale for spin lattice
relaxation is set by the {\em magnetocrystalline anisotropy energy}, which
is of the order of 100 $\mu $eV - 1 meV at surfaces, in thin
magnetic films or in hexagonal bulk crystals, rather
than by the Curie temperature or
by spin-orbit coupling or by electron-phonon
interaction (all being of the order of 30 - 50 meV). This energy scale comes
into play, since spin-lattice relaxation orginates from
the coupling of the spins to the {\em anisotropic} crystal field
fluctuations resulting from the phonons. These
fluctuations flip the spins to accomodate their thermal occupation to the
lattice temperature (or to a common equilibrium spin-lattice temperature).
Although magnetocrystalline anisotropy results from spin-orbit coupling,
its energy scale is typically
smaller at interfaces or in the bulk of noncubic
threedimensional solids by a factor of 100, since
spin-orbit coupling enters to second order (see appendix C). In cubic
bulk crystals the leading terms are of
fourth order thus resulting in a reduction factor of 10000. This argument holds
for both (i) the level shifts induced by spin-orbit coupling and
(ii) the occurrence and lifting of degeneracies at the Fermi
energy within a small portion of the Brillouin zone~\cite{moos}.
Our argumentation is still valid even for the particular case of Gd, where
the localized $f$-shell carries most of the magnetic moment while the
conduction electrons are responsible for the metallicity, since the
anisotropy of the magnetic moments involves the coupling of
localized and conduction electrons. The same holds for the
spin-lattice relaxation.
\\
\\
{\bf III. CONCLUSIONS}\\
In this work we presented a microscopic
theory for the spin-lattice relaxation time $\tau_{SL}$ in the metallic
rare earth ferromagnet Gd and found a value of 48 ps
in remarkably good agreement with experiment. Although
our theoretical estimate neglects all detailed features of electronic
structure,
phonon density of states, electronic correlations,
effects of electronic temperature, and the detailed form of the
transition matrix elements it already yields the correct value of $\tau_{SL}$.
Moreover, our theory clearly demonstrates the important relationship
between the static magnetocrystalline anisotropy energy and the dynamic
quantity $\tau_{SL}$, which is essential for magneto-optic recording
velocities.
Furthermore, our theory yields a good starting point for a detailed
electronic and nonequilibrium response theory of spin-lattice relaxation
in rare earth and transition metals (involving phonon-magnon coupling,
see appendix D). Thus it could overcome the restriction of previous ESR
theories to localized magnetic impurity spins (e. g. in insulating garnets).
For thin films and multilayers, it is of particular importance to calculate the
thickness dependence of this dynamical quantity, thus checking the range of
validity of the relation of $\tau_{SL}$ to magnetocrystalline anisotropy
and to the linear and nonlinear magneto-optical Kerr-effects, which involve
the complementary non-spin-flip effects of spin-orbit interaction.
It is a considerable theoretical challenge to investigate phonon-magnon
coupling in realistic itinerant systems.
{}From the experimental side,
additional thickness dependent and {\em spectroscopic} pump and probe
laser experiments as well as measurements of the ferromagnetic resonance (FMR
yielding collective spin-flip frequencies) are
required to tackle the important and complex problem
of spin-lattice relaxation in metallic ferromagnetic thin film media
and to bridge the gap between magnetic resonance experiments in the
frequency domain and optical real-time measurements.
In particular, it will be interesting to study the temperature dependence of
$\tau_{SL}$, thus discussing also low-temperature contributions to the
relaxation originating from direct or Orbach proceses.\\
Besides, it is of considerable interest
to search for faster spin-switching mechanisms using intense fs laser pulses
which may directly lead to a breakdown of magnetism via electron-electron
correlations and may therefore bypass the lattice thus reducing lattice
heating. It is to be
expected that more interesting results will be found
on the {\em femtosecond} time scale
which is now also accessible using Ti-sapphire lasers. Upon intense laser
excitation, the magnetic state may break down already within some fs
without the influence of the lattice and it is recovered within $\tau_{SL}$
which involves coupling of the spins to the
lattice via {\em anisotropic} crystal field fluctuations. In this case, the
spins are cooled by the lattice rather than heated as in the experiment
by Vaterlaus{\em et al.}, which requires a theoretical explanation.
These time scales should be optically accessible in metallic thin
film media in the near future.\\
\\
{\bf ACKNOWLEDGMENT}\\
We gratefully acknowledge stimulating discussions
with Prof. K. Baberschke, Dr. P. J. Jensen, and Dr. F. Meier.
\\
\\
\\
{\bf APPENDIX A: SPECIFIC HEAT OF PHONONS, SPINS, AND ELECTRONS}\\
In this appendix, we compare the specific heat of phonons, spins, and
electrons~\cite{ashcroft} in order to support our approximation that, at not
too low temperatures, a 10 ns laser pulse heats up mainly phonons.

The low-temperature specific heat of
{\em phonons} at constant volume is given within the Debye model by
\begin{equation}
c_{V}\;=\;\frac{12\pi^{4}}{5}nk_{B}(\frac{T}{\Theta_{D}})^{3}\;=\;
234\frac{T}{\Theta_{D}}^{3}nk_{B}.
\end{equation}
Here,
$\Theta_{D}$ denotes the Debye temperature, $k_{B}$ is Boltzmann's constant,
and $n$ is the number of lattice sites per unit volume.\\
The specific heat of the {\em spins} is given by
\begin{equation}
c_{H}\;=\;\frac{1}{3}\frac{N}{V}k_{B}J(J+1)(\frac{g\mu_{B}H}{k_{B}T})^{2}\;,
\end{equation}
where $J$ is the total angular momentum
and $g$ is the gyromagnetic ratio.\\
The specific heat of the {\em electrons} is
\begin{equation}
c_{V}\;=\;(\frac{\partial \mu}{\partial
T})\;=\;\frac{\pi^{2}}{3}k_{B}^{2}T\rho(\varepsilon_{F})\;=\;\frac{\pi^{2}}{2}
\frac{k_{B}T}{\varepsilon _{F}}nk_{B}.
\end{equation}
The spins start to dominate the phonons at temperatures $T$
\begin{equation}
T\;\leq \;T_{0} \;=\;(\frac{N}{N_{i}})^{\frac{1}{5}}
(\frac{g\mu_{B}H}{k_{B}T})^{\frac {2}{5}}\Theta_{D} \;\approx \;0.1\;\Theta_{D}
\end{equation}
for fields of about 1 Tesla and $N_{i}$ paramagnetic ions. The electrons
start to dominate the phonons at Temperatures $T$
\begin{equation}
T\;\leq \;T_{0} \;=\;0.145(\frac{Z\Theta_{D}}{T_{F}})^{\frac{1}{2}}\Theta_{D}
\;\approx \;0.01\;\Theta_{D},
\end{equation}
where $Z$ is the nominal valence. Typical Debye temperatures in ferromagnets
are
$\Theta_{D}$ = 420 K (Fe),
385 K (Co), 375 K (Ni), 152 K (Gd), 186 K (Dy).
This crude estimate shows already,
that, for ns laser pulses, phonons are dominant at not too low temperatures.\\
\\
{\bf APPENDIX B: DIRECT AND ORBACH PROCESSES}\\
In this appendix, we discuss the direct and Orbach relaxation processes
which may dominate the Raman contributions to spin-lattice relaxation
only at very low temperatures.\\
a) Direct process (Fig. 1):\\
This process consists of a
{\em spin-flip} and the {\em emission} of a phonon of frequency $\omega_{0}$.
The longitudinal relaxation rate is proportional to the temperature
\begin{equation}
\frac{1}{T_{1}}\;\sim \;T.
\end{equation}
The rate is also proportional to the number of phonons within a narrow
interval $\delta $ at the extreme low-frequency end of the phonon spectrum
\begin{equation}
\bar{p}_{0}(\delta)\;\approx \;\frac{k_{B}T}{\delta }.
\end{equation}
Thus, the direct process
is of importance only for very low temperatures ($T \; \ll\;\Theta_{D}$),
where the other processes become negligible.
The relaxation rate for the direct process depends on the magnetic field
(for magnetic impurities). It is proportional to $H^{4}$
for Kramers-doublets and proportional to $H^{2}$ for non-Kramers-doublets.\\
\\
b) Orbach process (Fig. 2):\\
This process~\cite{orbach} consists of a {\em spin-flip}, the {\em absorption}
of a phonon of frequency $\frac{\Delta}{\hbar}$, and the {\em emission}
of a phonon of frequency $\frac{\Delta}{\hbar}\;+\;\omega _{0}$. The
longitudinal relaxation rate for the Orbach process is approximately given by
\begin{equation}
\frac{1}{T_{1}}\;\sim \;e^{-\frac{\Delta_{1}}{k_{B}T}}.
\end{equation}
This rate corresponds to two high frequency cascades
and is proportional to the number of phonons in a narrow band at
\[
\Delta _{1}\;\approx {\rm crystal\;-\;field\;splitting}
\]
and does not depend on the magnetic field.
The Orbach process becomes important if the relation holds:
\begin{equation}
\frac{\Delta _{1}}{k _{B}}\;<\;\Theta _{D}.
\end{equation}
At higher temperatures, however, such as in the experiment by Vaterlaus {\em et
al.}, the Raman process should dominate the contributions originating from both
the direct and Orbach processes.\\
\\
{\bf APPENDIX C: MAGNETOCRYSTALLINE ANISOTROPY}\\
In this appendix, we give a simple estimate of magnetocrystalline anisotropy
in metals, which nevertheless contains most of the features
of a complete bandstructure calculation of
this quantity and already yields the correct order of magnitude. For
that purpose, we consider
a single, for simplicity parabolic,
but spin-orbit split band (Fig. 4). Hereby
we neglect the fact that parabolic bands
usually represent $s$ electrons which feel neither spin-orbit nor exchange
interaction. In addition, we neglect the magnetic dipole-dipole coupling
which favors in-plane magnetization in two
dimensions and is zero in the bulk of cubic or hexagonal crystals such as Gd.
It is in particular the spin-orbit induced magnetocrystalline anisotropy energy
which may (but does not necessarily have to) favor a perpendicular
easy axis in thin films and is therefore
of interest for high-density magnetic recording
(the time limit of which is related to $\tau_{SL}$).

We calculate now the
maximum energy gain from magnetocrystalline anisotropy in this model.
This gain originates
from the change of the band occupation up on spin-orbit induced lifting
of the band degeneracy at the Fermi level.
Electrons are transferred from one branch of the band to the other.
Assuming a Brillouin sphere in three dimensions one therefore
obtains an anisotropy energy of
\begin{equation}
E_{anisotropy}\;=\;\lambda_{s.o.}\times
\frac{4\pi k_{F}^{2}\Delta k}{\frac{4\pi}{3}k_{F}^{3}}
\;=\;\lambda_{s.o.}\times \frac{3\Delta k}{k_{F}}\;\;\;(3D),
\end{equation}
where
\begin{equation}
\frac{\Delta k}{k_{F}}\;=\;\frac{1}{2}\times \frac{\bar{k}}{k_{F}}
\end{equation}
is the number of states contributing to the change of the
electronic occupation. For typical values of the spin-orbit coupling constant
$\lambda_{s.o.}$ = 70 meV and the Fermi energy $E_{F}$ = 10 eV we find the
result
\begin{equation}
E_{anisotropy}\;=\;735\;\mu eV\;\;(3D).
\end{equation}
Thus, this crude model yields already important
insights in some of the microscopic features of magnetocrystalline anisotropy
which are confirmed by detailed calculations~\cite{moos}:
(i) The model gives the correct order of magnitude for
$E_{anisotropy}$ in films or non-cubic bulk crystals. The actual value for
Gd might be somewhat smaller in Gd but this would even improve the
agreement of $\tau_{SL}$
with experiment.
(ii) The model shows
that the magnetocystalline anisotropy
is smaller than the spin-orbit coupling constant
by two orders of magnitude since only a relatively
small portion of states close to the Fermi level may gain energy from
the spin-orbit induced lifting of degeneracies. For all other
states the upward and downward shifts of the lifted degeneracies cancel.
(iii) The model immediately yields that the anisotropy
energy resulting from the lifting of degeneracies
is proportional to the {\em square} of spin-orbit coupling, since
besides the explicit linear dependence on $\lambda_{s.o.}$
also the portion of contributing
states is linear in $\lambda_{s.o.}$. There is no azimuthal dependence on
spin-orbit coupling in this model in remarkable agreement
with the line degeneracies found in Fe monolayers~\cite{moos}. Thus, the energy
gain resulting from the {\em lifting of degeneracies}
close to the Fermi energy is of the same order of magnitude
as the magnetocrystalline anisotropy originating from {\em level shifts} far
below $E_{F}$, which can be obtained already in nondegenerate second-order
perturbation theory with respect to spin-orbit coupling.
(iv) The model explains why perpendicular
anisotropy may be favored in thin films: Due to the reduced
coordination number in these films, narrow bands of large density
of states close to the Fermi level (in ferromagnets) may occur which
can gain a suffciently large amount of magnetocrystalline anisotropy energy.
(v) The anisotropy at interfaces and the nonlinear magneto-optical Kerr-Effect
are closely related
via spin-orbit coupling although the latter results
from nonlinear optical excitations.
(vi) Interface hybridization of a ferromagnet with a
strong spin-orbit scatterer may yield large
anisotropies due to the reoccupation of many
states close to their common Fermi level.
(vii) Spin-orbit coupling does not
split exchange-split bands again. Thus, the diagonal part
\begin{equation}
L_{z}S_{z}
\end{equation}
just yields contrary level shifts of spin-up and
spin-down bands whereas the diagonal
contribution
\begin{equation}
L_{x}S_{x}\;+\;L_{y}S_{y}\;=\;\frac{1}{2}(L_{+}S_{-}\;+\;L_{-}S_{+})
\end{equation}
yields spin-flips. It is these spin-flips that contribute
to the spin-lattice relaxation time $\tau_{SL}$
which in this view describes the time required for adapting
the (temperature dependent) magneto-crystalline anisotropy to the lattice
temperature.
Thus, our model yields the correct order of magnitude and a change
of the direction of the magnetic moments.\\
\\
\newpage
\noindent
{\bf APPENDIX D: PHONON-MAGNON COUPLING}\\
If we want to conceive a similar theory for spin-lattice
relaxation in ferromagnetic transition metals
such as Fe we have to notice the following important difference:
In the rare earth metal Gd it is sufficient
to consider the localized $f$-electron spins carrying the magnetic moment of
7 $\mu_{B}$,
which gives rise to seven possible orientations
($m_l$ quantum numbers).
Thus, Raman processes may easily take place.
In the transition metal Fe, however, the
$d$ electrons are itinerant (delocalized), have to be described within
the band picture, and have only
two orientations (spin-up and spin-down).
Therefore collective and quantized magnetic
excitations (magnons) have to be allowed for the spin-lattice
relaxation (Fig. 5).
The transition Hamiltonian then describes the phonon-magnon coupling
using boson creation and annihiliation
operators $a^{(\dagger )}_{{\bf k}}$ for magnons and
$b^{(\dagger )}_{{\bf k}}$ for phonons
\begin{equation}
H_{ph-mag}({\bf k})\;=\;D_{{\bf k}}[b_{{\bf k}}a^{\dagger }_{{\bf k}}\;+\;h.c.]
\end{equation}
with~\cite{balucani}
\begin{equation}
D_{{\bf k}}\;=\;[\frac{3}{3\sqrt{2}}\frac{D}{\varepsilon }F]\sqrt{2S}(2s-1)
\sqrt{\frac{\hbar \mid {\bf k}\mid}{2M\bar{v}_{s}}}.
\end{equation}
Here, $D$ is some coupling strength, $F$
some crystal field parameter, $S$ the spin, $M$ the effective
mass, and $\overline{v} _{s}$ the averaged speed of
sound.\\
\\
The interaction Hamiltonian $H_{ph-mag}({\bf k})$ can then be inserted
in the full phonon-magnon Hamiltonian~\cite{kittel}
\begin{equation}
H\;=\;\sum_{{\bf k}}[\omega^{m}_{{\bf k}}a^{\dagger}_{{\bf k}}a_{{\bf k}}
\;+\;\omega^{p}_{{\bf k}}b^{\dagger}_{{\bf k}}b_{{\bf k}}
\;+\;H_{ph-mag}({\bf k})]
\end{equation}
which is easily solved by applying the unitary transformation
\begin{eqnarray}
a^{(\dagger )}_{{\bf k}}\;&=&\;A^{(\dagger )}_{{\bf k}}\cos\Theta_{{\bf k}}\;
+\;B^{(\dagger )}_{{\bf k}}\sin\Theta_{{\bf k}}\nonumber\\
b^{(\dagger )}_{{\bf k}}\;&=&\;B^{(\dagger )}_{{\bf k}}\cos\Theta_{{\bf k}}\;
-\;A^{(\dagger )}_{{\bf k}}\sin\Theta_{{\bf k}}
\end{eqnarray}
to yield
\begin{eqnarray}
\omega_{A}\;&=&\;\omega^{m}_{{\bf k}}\cos^{2}\Theta_{{\bf k}}
\;+\;\omega^{p}_{{\bf k}}\sin^{2}\Theta_{{\bf k}}\;
-\;2D_{{\bf k}}\cos\Theta_{{\bf k}}\sin\Theta_{{\bf k}}\nonumber\\
\omega_{A}\;&=&\;\omega^{m}_{{\bf k}}\cos^{2}\Theta_{{\bf k}}
\;+\;\omega^{p}_{{\bf k}}\sin^{2}\Theta_{{\bf k}}\;
+\;2D_{{\bf k}}\cos\Theta_{{\bf k}}\sin\Theta_{{\bf k}}.
\end{eqnarray}
$\Theta_{{\bf k}}$ is given by
\begin{equation}
\tan 2\Theta_{{\bf k}}\;=\;\frac{2D_{{\bf k}}}{\omega^{p}_{{\bf k}}-
\omega^{m}_{{\bf k}}}.
\end{equation}
This then completes
the formal solution of the phonon-magnon problem.\\
\\

\newpage
\begin{figure}
\caption{(a) Direct process, (b) Orbach process, and (c) Raman process.}
\label{fig1}
\end{figure}
\begin{figure}
\caption{Microscopic model for magnetocrystalline anisotropy energy.}
\label{fig2}
\end{figure}
\begin{figure}
\caption{Spin-lattice relaxation in transition metals.}
\label{fig3}
\end{figure}
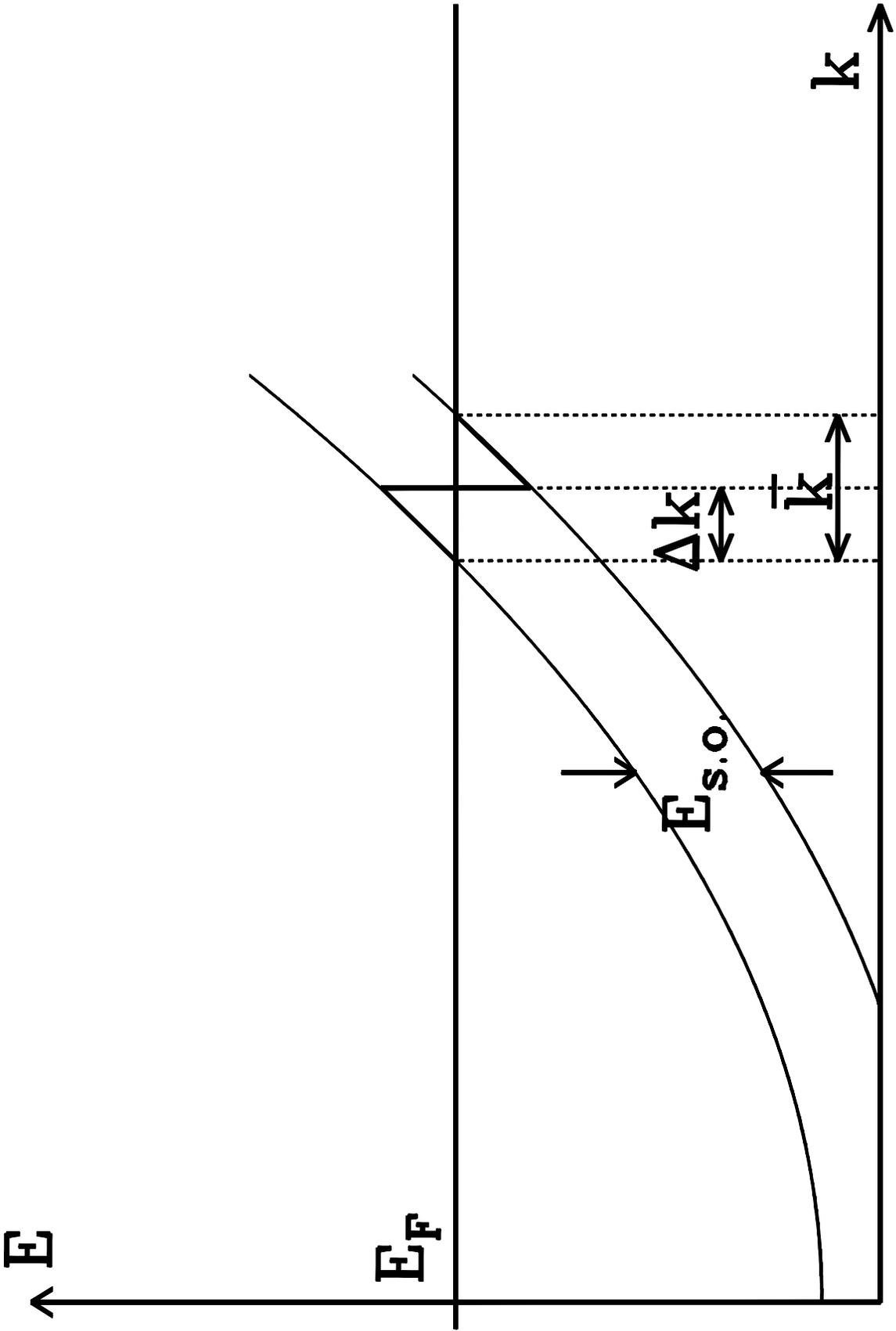
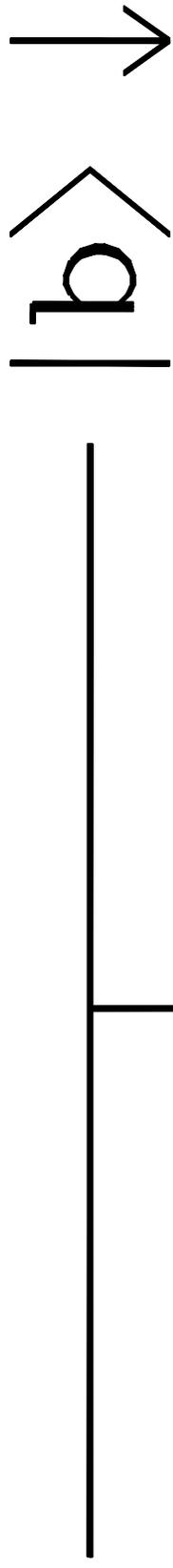
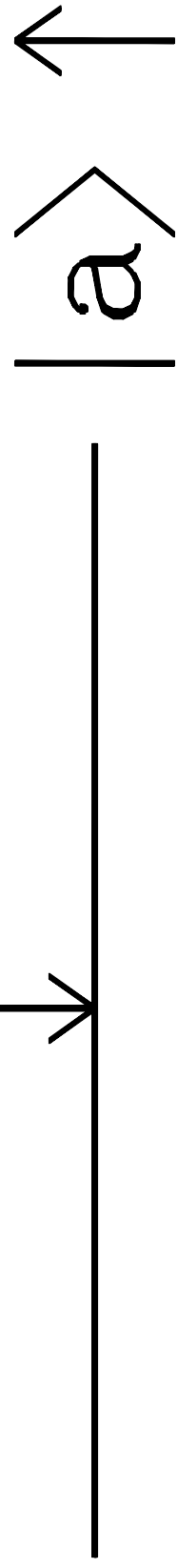
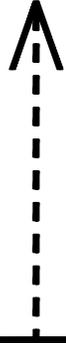
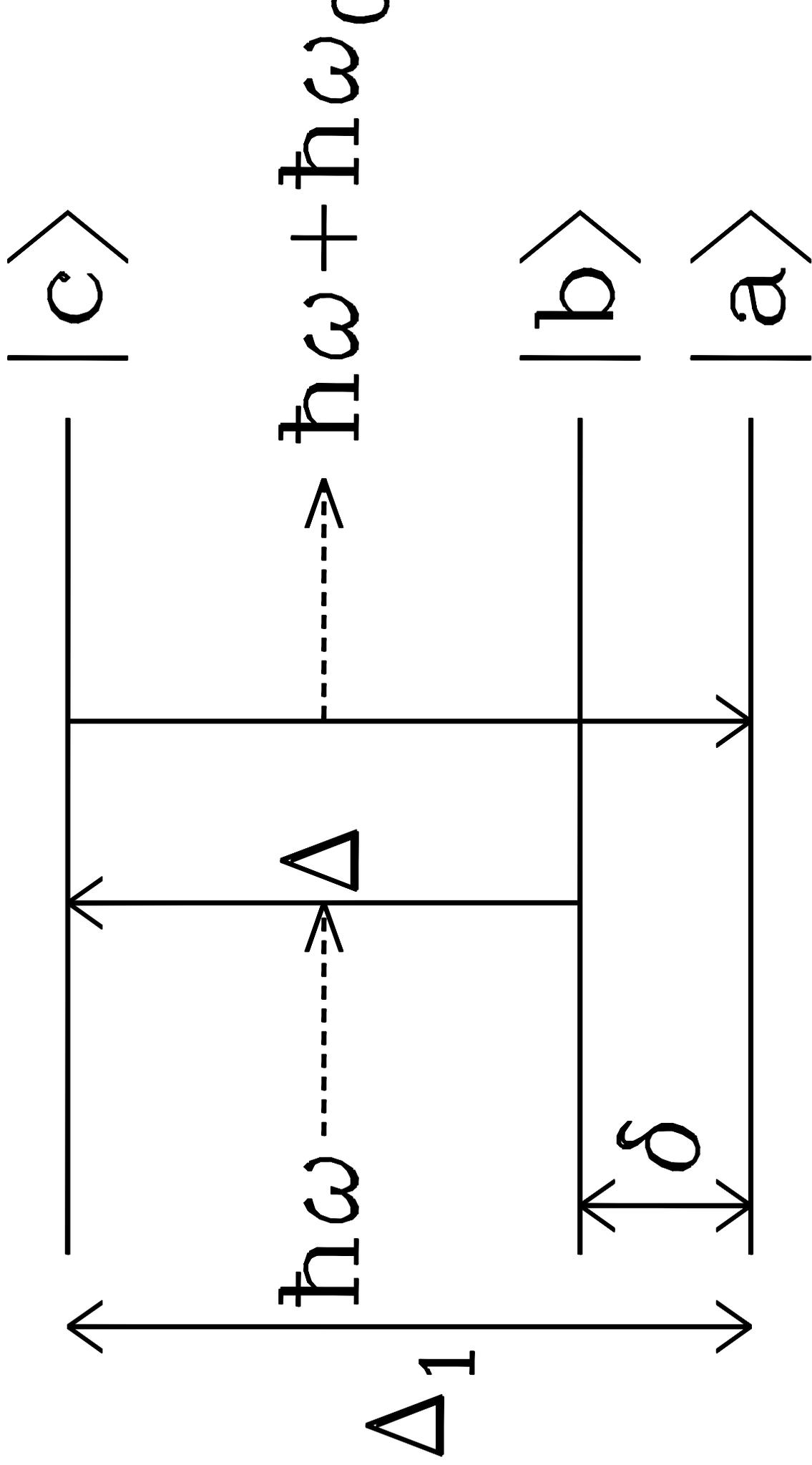
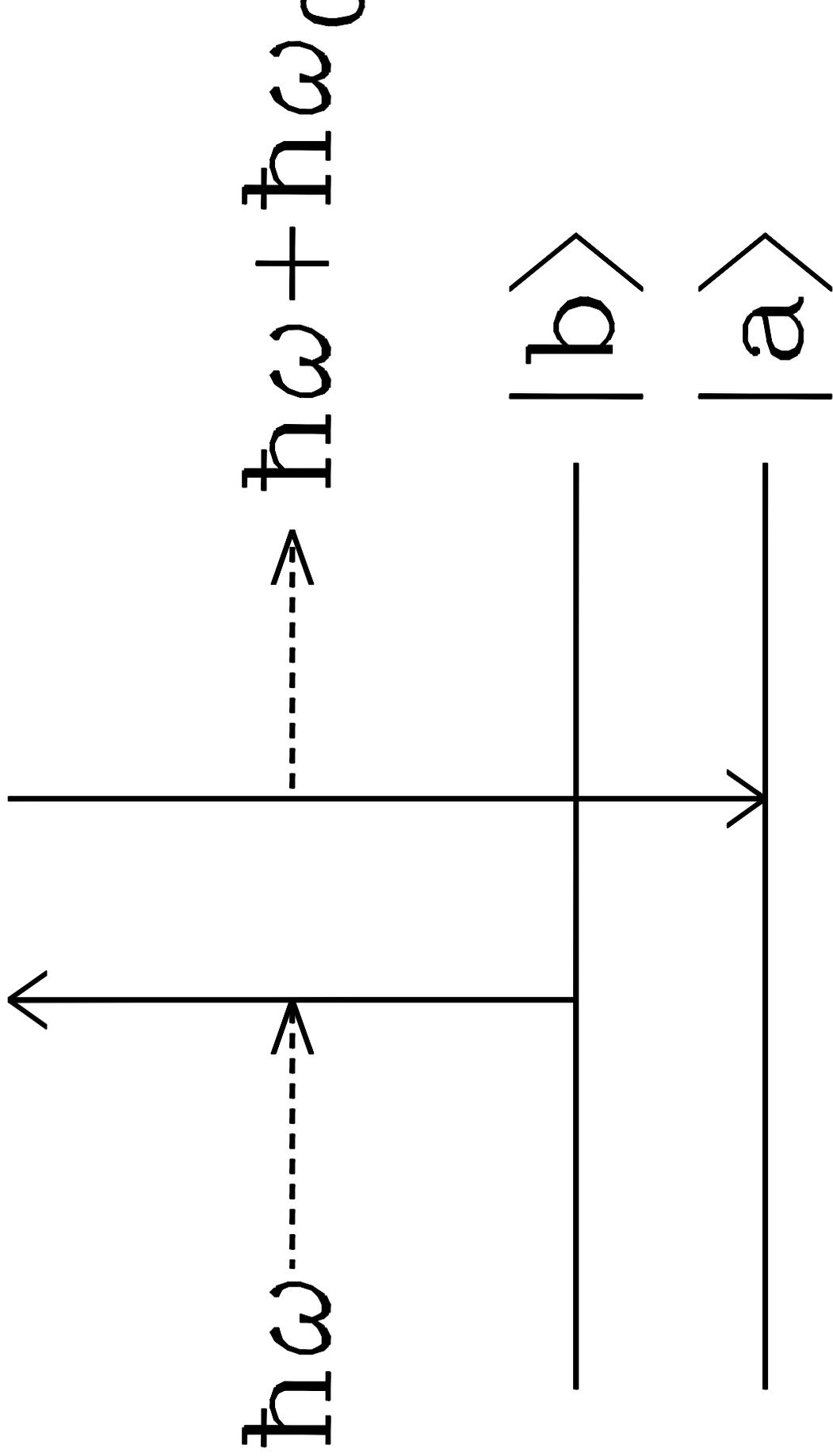
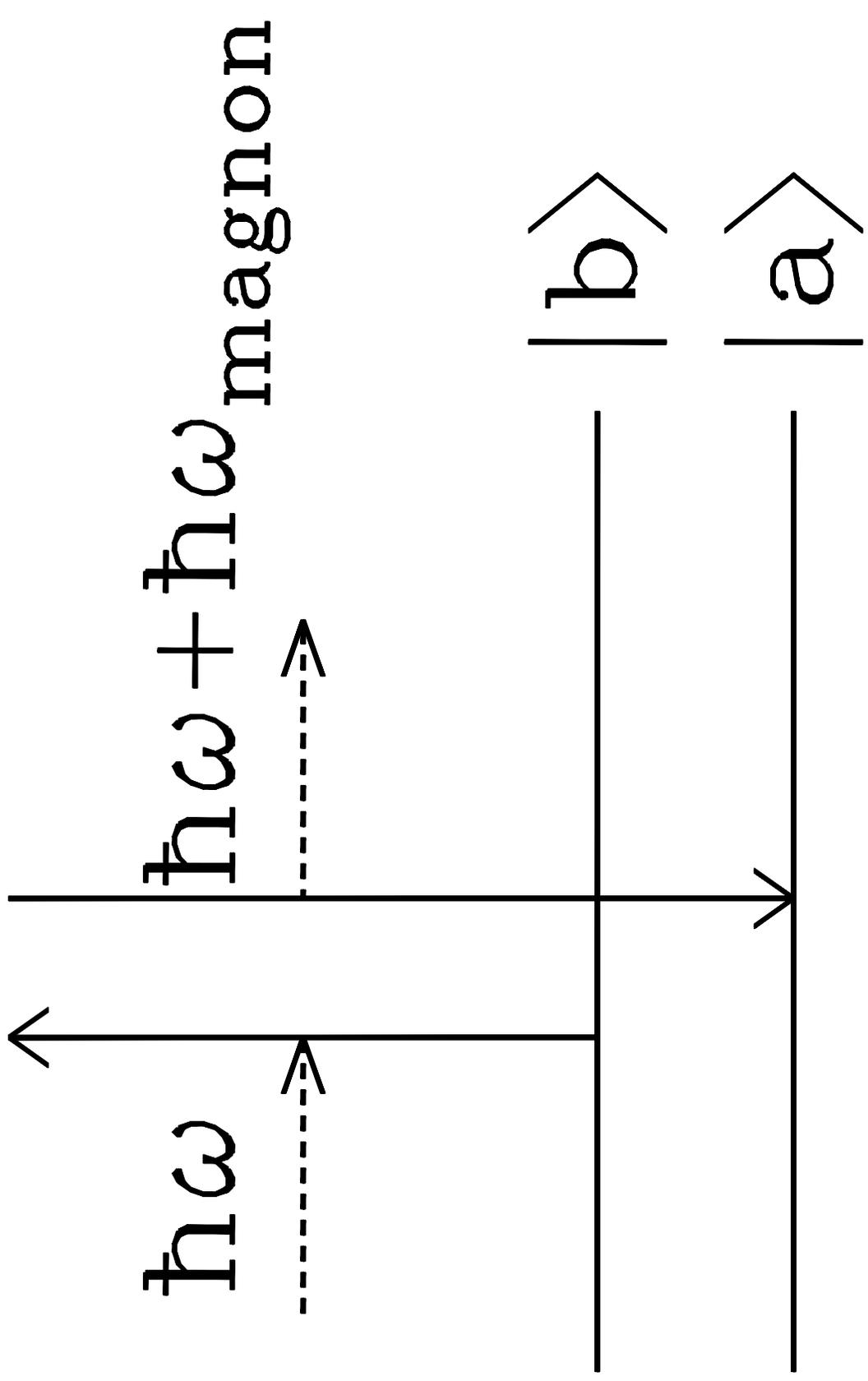
\end{document}